\documentclass[fleqn,usenatbib,useAMS]{mnras}

\usepackage{graphicx} 
\usepackage{amsmath} 
\usepackage{amssymb} 
\usepackage{multicol} 
\usepackage{bm} 
\usepackage{pdflscape} 
\usepackage{graphicx,xcolor,hyperref}

\usepackage{xcolor}

\usepackage[T1]{fontenc}
\usepackage{ae,aecompl}

\newcommand{\hi}{\textsc{Hi}}

\newcommand{\be}{\begin{equation}}
\newcommand{\ee}{\end{equation}}
\newcommand{\bea}{\begin{eqnarray}}
\newcommand{\eea}{\end{eqnarray}}
\newcommand{\bfig}{\begin{figure}}
\newcommand{\efig}{\end{figure}}
\newcommand{\nn}{\nonumber}

\newcommand{\ylm}[1]{Y_{\ell m}( #1)}
\newcommand{\ylmc}[1]{Y^*_{\ell m}( #1)}
\newcommand{\almi}[1]{a_{\ell m,#1}}
\newcommand{\almpi}[1]{a_{\ell' m',#1}}

\def\alm{a_{\ell m}}

\def\n{\hat{\boldsymbol{n}}}
\def\l{\left(}
\def\r{\right)}
\def\alm{a_{\ell m}}

\def\O{\Omega}
\def\don{{\rm d}\O_{\hat n}}

\def\fnl{{f_{\rm NL}}}
\def\egr{{\epsilon_{\rm GR}}}
\def\d{\mathrm{d}}

\def\l{\left(}
\def\r{\right)}
\def\D{\Delta}
\def\H{{\mathcal H}}
\def\p{\partial}
\def\fsky{f_{\rm sky}}
\def\z{(z)}

\def\hi{\textsc{Hi} }
\def\e{\textsc{e}}

\title[Ultra-large scale effects and \hi IM foregrounds]{Measuring ultra-large scale effects in the presence of 21cm intensity mapping foregrounds}

\author[J.\ Fonseca \& M.\ Liguori]{Jos\'e Fonseca$^{1,2,3,4}$\thanks{josecarlos.s.fonseca@gmail.com} and Michele Liguori$^{1,2}$\\
$^1$Dipartimento di Fisica ``G. Galilei'', Universit\`a degli Studi di Padova, Via Marzolo 8, 35131 Padova, Italy\\
$^2$INFN -- Istituto Nazionale di Fisica Nucleare, Sezione di Padova, Via Marzolo 8, 35131 Padova, Italy\\
$^3$School of Physics \& Astronomy, Queen Mary University of London, London E1 4NS, UK\\
$^4$ Department of Physics \& Astronomy, University of the Western Cape, Cape Town 7535, South Africa\\ 
}

\date{}

\pubyear{2020}

\begin{document}
\label{firstpage}
\pagerange{\pageref{firstpage}--\pageref{lastpage}}
\maketitle

\begin{abstract}
\hi intensity mapping will provide maps of the large-scale distribution of neutral hydrogen (\textsc{Hi}) in the universe. These are prime candidates to be used to constrain primordial non-Gaussianity using the Large Scale Structure of the Universe as well as to provide further tests of Einstein's theory of Gravity (GR). But \hi maps are contaminated by foregrounds,  which can be several orders of magnitude above the cosmological signal. Here we quantify how degenerated are the large-scale effects ($\fnl$ and GR effects) with the residual foregrounds. We conclude that a joint analysis does not provide a catastrophic degradation of constraints and provides a framework to determine the marginal errors of large scale-effects in the presence of foregrounds. Similarly, we conclude that the macroscopical properties of the foregrounds can be measured with high precision. Notwithstanding, such results are highly dependent on accurate forward modelling of the foregrounds, which incorrectly done catastrophically bias the best fit values of cosmological parameters, foreground parameterisations, and large-scale effects.

\end{abstract}





\section{Introduction}
\label{sec:intro}

Our understanding of the standard cosmological model has incrementally improved in the past decades. Some of the main questions cosmologists have been trying to answer concern the nature of Dark Energy and Dark Matter, with uncertainties on their physical properties becoming constrained to a few percent level. This $\Lambda$CDM concordance model is backed by different types of observations (despite some inconsistencies among them) such as the Cosmic Microwave Background (CMB) \citep{Aghanim:2018eyx}, Supernovae Type Ia \citep[for a recent measurement see][]{Riess:2018byc}, and the Large Scale Structure (LSS) of the Universe \citep[see for example the latest SDSS results][]{2021MNRAS.501.5616D}. 

Beyond the nature of the dark components of the universe, understanding the very early Universe and the seeds of the LSS still remains interesting open questions. Although the latest Planck results \citep{Aghanim:2018eyx} give good constraints on the amplitude and spectral index of the primordial curvature fluctuation, these alone are not enough to distinguish between inflationary models \citep{Akrami:2018odb}. All of these are based on the description that the primordial energy density field is a Gaussian random field sourced by quantum fields during inflation \citep[see][for a review]{Bassett:2005xm}. Standard single field inflationary models predict the density field to be nearly Gaussian \citep{Maldacena:2002vr} but a plethora of other models predict deviations from the Gaussian assumption either due to the presence of other fields \citep[see for example the Curvaton scenario][]{Lyth:2001nq}, inhomogeneous reheating at the end of inflation \citep{Dvali:2003ar}, or non-standard inflationary set-ups. For a review on primordial non-Gaussianity (PNG) from Inflation refer to \citet{Bartolo:2004if}. The latest Planck results already put bounds on primordial non-Gaussian parameters \citep{Akrami:2019izv} giving the state-of-the-art measurement of local-type primordial non-Gaussianity $f_{\rm NL}^{\rm local} = -0.9 \pm 5.1$. Note that CMB and LSS use different normalisations for PNG $\fnl^{\rm LSS}\simeq 1.3 \fnl^{\rm CMB}$ \citep{Camera:2014bwa}. Local $\fnl$ measures the leading order expansion of the gravitational potential and induces a scale-dependent correction to the bias of Dark Matter tracers \citep{Matarrese:2008nc,Dalal:2007cu}.   

Such non-Gaussian correction to the bias has a $1/k^2$ dependence, being only relevant on very large scales. Similarly, on these scales, other observational effects - so-called ``GR effects" - affect how we measure the density fluctuations of biased tracers in the LSS \citep{Yoo:2012se,Challinor:2011bk,Bonvin:2011bg}. In fact, primordial non-Gaussianity and GR effects on the past light cone are degenerate, and neglecting such effects will bias estimates of $\fnl$ using the LSS \citep{Camera:2014bwa}. Hence, any analyses intended to measure $\fnl$ has to include the GR effects as they also scale with $1/k$ and $1/k^2$. One can immediately see that we require surveys that will cover large volumes of the universe to see effects on super-horizon scales. 

Future galaxy surveys such as Euclid \citep{Amendola:2016saw} and the Vera C. Rubin Observatory \citep[former LSST][]{Abate:2012za} will observe and create catalogues of galaxies in large areas of the sky as well as deeper in redshift. While galaxy surveys are a well establish probe of the LSS, there are other tracers of dark matter. A promising way is to use the emission of 21cm photons of the hyperfine transition of neutral Hydrogen (\textsc{Hi}) and produce maps of the intensity of the signal \citep{Chang:2007xk,2012A&A...540A.129A}. After the Epoch of Reionisation, neutral Hydrogen remains only in galaxies while the inter-galactic medium becomes ionised. Hence, measuring the statistical distribution of \hi intensity we are probing underlying dark matter distribution. Although a weak line, \hi is by far the largest component of the baryonic matter making such a signal visible. In \hi intensity mapping (IM) one does not detect individual galaxies, instead one produces CMB-like maps of the \hi emission. The planned Square Kilometre Array (SKA) will allow \hi IM tomography in single dish mode \citep{Bacon:2018dui}, with its precursor MeerKAT \citep{2017arXiv170906099S} already operational. \hi IM ideal to study $\fnl$ as it permits fast scans of the sky with high redshift resolution covering a wide range of frequencies \citep{Camera:2013kpa,Alonso:2015uua}.

Despite the great potential of \hi IM for cosmology, obtaining the clean \hi signal is a challenge due to the presence of galactic and extragalactic foregrounds that contaminate the cosmological signal. In fact, they are several orders of magnitude above the cosmological signal \citep{Santos:2004ju} and pose a challenge if one wants to do cosmology with \hi IM. But there are strategies to deal with the presence of foregrounds which can divide into two main approaches. One such strategy is \emph{foreground avoidance} which, roughly speaking, sets to find an observational window of the power spectrum in $k_{||}-k_\perp$ space where the foregrounds do not affect substantially the estimates of the power spectrum \citep[see for example][]{Shaw:2014khi}. The wedge cuts typically renders scales $k_{||}\lesssim 0.01$Mpc$^{-1}$ inaccessible due to the smoothness in frequency of the foregrounds. Such an approach is therefore unsuitable for primordial non-Gaussianity studies with \hi IM as no large scales are present in the observational window. The other approach, and extensively studied in the literature, is to do \emph{foreground cleaning}. Most of the foreground removal methods take advantage of their spectral smoothness, while the cosmological \hi signal is expected to fluctuate in the line-of-sight. Earlier attempts to subtract foregrounds used polynomials of the logarithm of the frequency \citep{Bowman:2008mk} to capture their contribution to the observed signal. These parametric approaches are not the preferred way of dealing with foregrounds as they put a strong prior to their frequency structure. A much-preferred way of performing foreground removal is by using blind methods that make little assumptions about the foregrounds, except that they are smooth in frequency. These are very similar to what has been done with CMB experiments. Blind methods have been extensively used to recover the LSS power spectrum of \hi IM including using Independent Component Analyses (ICA) and principal component analyses (PCA) \citep{Alonso:2014dhk,Zhang:2015mga,Asorey:2020mxs}, GMCA (Generalised Morphological Component Analysis) and other sparse methods in pixel space \citep{Carucci_2020,2021MNRAS.tmp..875C}. These methods are able to remove the foreground contamination up to certain scales. Generically they remove the mean of the \hi signal and suppress the power on large scales where the signal we aim to detect resides. One can try to reconstruct the power on large scales using transfer functions \citep{Switzer:2015ria,Witzemann:2018cdx} but these fundamentally depend on the input calibration, potentially leading to biased results if the transfer function has been calibrated with the wrong cosmology or with an incomplete setup \citep[for example]{Wolz:2013wna}. Therefore to study the effect of primordial non-Gaussianity in the \hi power spectrum using foreground subtraction, one needs to understand how the blind methods suppress the large scale power \citep[see for example][]{2020MNRAS.499.4054C}.

Here we take a different approach, neither we subtract nor we avoid foregrounds, and try to understand if we can still draw any conclusion of the cosmological model in their presence. Our work is similar to \citet{Santos:2006fp} although we focus on $\fnl$ and GR effects and in the post-EoR universe. As \citet{Camera:2019iwy} we look at the degeneracies between $\fnl$ and astrophysics. While they focused on the HI halo model we focus on the effects of Galactic astrophysics instead. We are also within the same lines of \citet{Zaldarriaga:2003du} but we consider the foregrounds to be well correlated in frequency and use \citet{Santos:2004ju} to take into account the correlation length of the foregrounds. Instead of just considering the spectral smoothness of the foregrounds, we also want to see if we can learn anything about their angular structure. Therefore we will forward model the foregrounds power spectrum and jointly understand their physical properties together with cosmological parameters in the spirit of \citet{Switzer:2018tel} and \citet{Sims:2019iyz}. We will use the standard Fisher formalism to forecast how well one can learn physical properties of the foregrounds and the cosmology, and assess how constraints on $\fnl$ and the GR effects are degraded once we marginalise the foreground model parameters. Bear in mind that when we speak about foregrounds, in practice we refer to residual foregrounds, as any calibration method includes models for galactic emission \citep{2020arXiv201113789W}. One should note that we use simplistic toy models for astrophysical foregrounds, but our main goal here is to understand if such an approach is at all feasible in practice. A more elaborated model should take into account previous modelling and mappings of the sky, such as of \citet{deOliveiraCosta:2008pb} although these do not break down into radio emission components. We do not include other systematics that will affect the result, like polarisation leakage \citep{Spinelli:2019oqm,2021MNRAS.tmp..875C}, 1/f noise \citep{Harper:2017gln,2021MNRAS.501.4344L}, beam asymmetries \citep{2021MNRAS.502.2970A}, RFI and others. We will consider the SKA1-MID (both bands) as our experimental setting, although any single dish mode setup with sufficiently low noise levels should provide similar results. Having a parametric model for the different foreground components seems a straight forward approach, but it may be a strong prior imposition on the foreground physics. Hence, we will try to understand how incorrect modelling of the foregrounds may affect our estimates. This can be catastrophic and introduce severe bias on the cosmological parameter estimation. The degeneracies between foregrounds and PNG have been started to be studied recently by \citet{2020MNRAS.499.4054C}. Although we try to understand the same degeneracy, we do not perform foreground cleaning and we work in harmonic space. We also use the Fisher matrix instead of a more correct MCMC as the Fisher formalism suffice for our set goal. 

Although we wanted to focus on the large scale effects, our work also looks at the same questions posed by \citet{Liu:2011aa} on what can be learnt about the foreground physics. While they conclude that there should be large degeneracies between the handful of parameters describing the physics of the foregrounds we find the opposite. We do not consider the same parameters and we look at summary statistics of the maps instead of studying how to decompose the different contributions at the pixel level. In the conclusion, we will discuss more in detail where this difference may come from. 

This paper is organised as follows: in \S \ref{sec:maps} we review the expected observed HI IM signal and all its components; in \S \ref{sec:methods} we review the Fisher matrix and the different approaches we will take; in \S \ref{sec:results} we present our forecasts for how well we can measure large scale effects as well as understand the foregrounds; in \S \ref{sec:bias_results} we gain insight of how biases we are when we have incorrect forward models; and we conclude \S \ref{sec:conclusions}.

\section{Recap on observed maps of intensity}
\label{sec:maps}

Generically speaking one can think of two types of intensity mapping experiments: one where detectors/receivers work as interferometers and immediately output the Fourier transform of the sky; the other where dishes, CCDs, bolometers, or thermometers measure temperature or intensity or brightness of the sky in a \emph{voxel} centred at a direction $\n$ and frequency $\nu_i$. In this paper, we will focus on the second type of experiment. Notwithstanding, the approach taken in this paper is, in principle, also applicable to interferometric experiments. Let us assume that after calibration and map making, the IM experiment provides data points ${\cal M}(\nu_i,\n)$ on the celestial sphere, where ${\cal M}$ can be temperature, intensity, or brightness, for example. As we will be focusing on radio experiments ${\cal M}$ will be a temperature map, although the framework can be general. In general, the observed map will contain contributions from the signal ${\cal S}$ we wish to detect as well as random noise ${\cal N}$ from the experiment itself and other foreground/background contaminants ${\cal F}$ in the sky. For \hi IM we only focus on foregrounds. These can be of several types: cosmological, galactic, satellites, atmospherical effects, etc. For this paper, we focus on galactic and extragalactic foregrounds as they are the dominant contribution to the signal. In addition, there are toy models we can explore and have been used extensively in the literature to simulate the foregrounds. On the other hand, satellites (and other transient effects) are harder to model and only affect some pointings during their passage through the dishes' field-of-view (FoV). Broadly speaking, atmospheric effects can be dealt with at the calibration level \citep{2020arXiv201113789W}. Our observed maps can then be expressed as 
\be \label{eq:map}
{\cal M}(\nu_i,\n)={\cal S}(\nu_i,\n)+{\cal F}(\nu_i,\n)+{\cal N}(\nu_i,\n)\,.
\ee
In cosmology, it is the statistical structure of density fluctuations that carries the information we wish to measure, therefore we work with contrast maps in frequency shells 
\be
\Delta {\cal M}(\nu_i,\n)\equiv {\cal M}(\nu_i,\n)-\bar{\cal M}(\nu_i) \,,
\ee
where $\bar {\cal M}$ is the sky average $\nu_i$ map. We then expand the contrast map into a spherical harmonic basis $\ylm{\n}$ as
\be
\Delta {\cal M}(\nu_i, \n)= \sum^{\infty}_{\ell = 0} \sum^{\ell}_{m = -\ell} \alm(\nu_i) \ylm{\n}\,,
\ee
where the $\alm$ coefficients only depend on the frequency bin. One can then use the orthonormal properties of the spherical harmonics to compute the $\alm$, which are given by
\be \label{eq:alm}
\alm(\nu_i)=\int \don\ \Delta{\cal M}(\nu_i,\n)\ \ylmc{\n}\,.
\ee

All relevant information about the cosmological signal (and the noise) is as usual captured by the power spectrum
\be
\Big\langle a_{\ell m}(\nu_i)\ a^{*}_{\ell' m'}(\nu_j)\Big\rangle = \delta_{\ell \ell'}\ \delta_{m m'}\ C_\ell\l \nu_i,\nu_j\r\,,
\ee
Note that the $C_\ell$ of the experimental noise and the signal have a different interpretation than the foregrounds one. While $\alm^{\cal S}$ is drawn from a Gaussian distribution centred at zero and variance $C_\ell^{\cal S}$ (same for the instrumental noise), the $\alm^{\cal F}$ and $C_\ell^{\cal F}$ represent the angular structure of the foregrounds which does not result from any intrinsic probability distribution. There is a caveat, this may not hold for extragalactic foregrounds as they may be indeed a realisation of the underlying density field. We will assume they have a fixed angular structure as galactic foregrounds do. From the map given by Eq. \ref{eq:map} we get
\be \label{eq:cl_map}
C_\ell^{\cal M}\l \nu_i,\nu_j\r=C_\ell^{\cal S}\l \nu_i,\nu_j\r+C_\ell^{\cal F}\l \nu_i,\nu_j\r+C^{\cal N}_\ell\l \nu_i,\nu_j\r \,,
\ee
where we considered each component uncorrelated from each other. This is undoubtedly true for the experimental noise. We also don't expect correlations between the cosmological signal and galactic foregrounds. Extragalactic point sources may indeed correlate with the cosmological signal although we neglect it for now. We expect correlations with extragalactic free-free emission to be negligible as it mainly comes from the intergalactic medium \citep{Santos:2004ju}.

From now on let us suppress the frequency dependence to simplify notation and use the indices $i,j,p,q$ to refer to frequency/redshift bins. We will therefore follow the notation $\alm \l\nu_i\r=\almi{i}$ and $C_\ell\l \nu_i,\nu_j\r=C_{\ell,ij}$. Note that the indices $\ell,m$ will always refer to the spherical harmonic decomposition multipoles. Our estimator of the angular power spectrum, in the full sky regime, is given by
\be\label{eq:cl_estim}
\hat C_{\ell,ij} =\frac{1}{2\ell+1}\sum_{m=-\ell}^{m=\ell} \frac12\Big[\almi{i}\ \almi{j}^*+\almi{j}\ \almi{i}^*\Big]\,.
\ee
In Appendix \ref{apx:cov_cl} we revise the covariance of the angular power spectrum estimator which is given by
\bea \label{eq:cov}
\l\Gamma_{\ell,\ell'}\r_{ij,pq}^{\rm full\ sky}&=&Cov\left[\hat C^{\cal M}_{\ell,ij}, \hat C^{\cal M}_{\ell',pq}\right]\nn\\
&=&\frac{\delta_{\ell,\ell'}}{2\ell+1} \l C^{\cal G}_{\ell,ip} C^{\cal G}_{\ell,jq} +C^{\cal G}_{\ell,iq} C^{\cal G}_{\ell,jp} \r \label{eq:cov}\,,
\eea
where we defined the ``Gaussian" $\cal G$ part of the map as
\be
C^{\cal G}_{\ell,ij} \equiv C_{\ell,ij}^{\cal S}+C^{\cal N}_{\ell,ij}\,.
\ee
The fact that the foregrounds do not contribute to the covariance is not unexpected. In the absence of cosmology and if we wanted to fit some given model of the angular and frequency structure of the foregrounds then the uncertainty budget would uniquely come from the instrument one uses. 
Nevertheless we will consider a modification of the covariance to mimic modelling uncertainties in the foregrounds. 

Any realistic survey covers a limited area in the sky. This implies that some $m$ will not be available to Eq. \ref{eq:cl_estim}. We can approximate the cut-sky covariance as
\be \label{eq:cov_fsky}
\l\Gamma^{\rm cut\ sky}_{\ell,\ell'}\r_{ij,mn}=\frac{\l\Gamma^{\rm full\ sky}_{\ell,\ell'}\r_{ij,mn}}{f_{\rm sky}}\,.
\ee
Note as well that we do not bin in $\ell$, otherwise we would need to further divide the covariance by $\Delta\ell$. In a realistic setting one always bins in $\ell$ (which is fundamentally determined by the survey mask) but this will not alter the conclusions of the paper.  
 
We now only need to model each component of Eq. \ref{eq:map} including large scale effects that are degenerate with the foregrounds. We will use interchangeably redshift and frequency. While observationally it is natural to use frequency, theoretically one usually speaks in term of redshift $z$. In HI IM (or in IM in general) frequency and redshift are equivalent since the signal comes from a given emission line. In our case 
\be
z+1=\frac{1.4{\rm GHz}}{\nu^{\rm obs}}\,
\ee
where $\nu_{\rm obs}$ is the observed frequency. Note however that redshift averages are not equivalent to frequency averages. Here we will only use narrow frequency/redshift bins with top-hat window functions where both averages are approximately equal. Also note that in a perturbed universe the observed redshift does not correspond to the true redshift of the source, especially due to peculiar velocities \citep{Kaiser:1987qv}.

\subsection{The observed cosmological signal}
\label{sec:effect_fnl_GR}

The observed angular power spectrum will depend on the true (theoretical) angular power spectrum convolved by the optics of the IM experiment and binned and weighted in frequency. We consider a radio observatory like the SKA1 which is composed of dishes. To first order, one can approximate the beam of the telescope as Gaussian, which in harmonic space becomes 
\be
B_\ell(z_i)=\exp\left\{\frac{-\ell(\ell+1)\theta_{FWHM}^2(z_i)}{16\ln 2}\right\}\,,
\ee
where the angular resolution is
\be
\theta_{FWHM}(z_i)=1.22 \frac{\lambda_{\rm HI}(1+z_i)}{D_{dish}}\,,
\ee
and $D_{dish}$ is the diameter of the dishes. Then the observed angular power spectrum is given by
\be
C^{\hi, obs}_\ell \!\l z_i,z_j\r=B_\ell(z_i) B_\ell(z_j) C^{ W_\hi}_\ell \!\l z_i,z_j\r
\ee
where $C^{W}_\ell$ takes into account the selection and source distribution functions and is given by \citep{Challinor:2011bk}
\bea \label{eq:cl_W}
C^{W_\hi}_\ell \!\l z_i,z_j\r=4\pi\!\!\int\!\!\d \ln k\,\D_\ell^{W_\hi}\!\l z_i,k\r \D_\ell^{W_\hi}\!\l z_j,k\r \mathcal P\!\l k\r \!\,.
\eea
The weighting does not happen at the power spectrum level itself, it is the transfer functions that are weighted by how we select the data and the distribution of sources. Therefore 
\bea \label{eq:transf}
\D_\ell^{W_\hi}\l z_i,k\r=\int\!\! \d z\,T^{\hi}(z)W(z_i,z)\D^\hi_\ell(z,k).
\eea
In the case of \hi IM, the distribution of sources is just the HI temperature $T^{\hi}$. The window function $W$ is given by how we bin and weight the data in frequency and is a normalised probability distribution function such that $\int\!\d z\,W(z_i,z)=1$. The transfer function $\D_\ell^{\hi}$ includes the underlying matter density fluctuations as well as Redshift Space Distortions (RSD) and general relativistic effects that alter the apparent density field. Eq. \ref{eq:cl_W} also relates the observed angular power spectrum with the dimensionless primordial curvature perturbation power spectrum
\bea
\mathcal P (k)=A_s\l \frac k{k_0}\r^{n_s-1}.
\eea
Here the pivot scale is $k_0=0.05\,$Mpc$^{-1}$, $A_s$ is the amplitude and $n_s$ is the spectral index. 

In the case of $\hi$ intensity mapping the expression of the transfer function in Newtonian gauge is given by \citep{Hall:2012wd}
\bea
\D^{\hi}_\ell(k)&=&\left[ \l b^{\hi}+\Delta b^{\hi}(k)\r \delta^{\rm s}_k\right] j_\ell\l k\chi\r +\frac {k v_k}{\H}j_\ell''(k\chi)\nn\\
&&{}+\epsilon_{GR}\Bigg\{\left[ \psi_k+\frac{\phi_k'}{\H}+\l b^{\hi}_e-3\r\frac {\H v_k}{k}\right] j_\ell\l k\chi\r\nn\\
&&{}+\l2-b^{\hi}_e+\frac{\H'}{\H^2}\r
 \bigg[ v_k j_\ell'(k\chi) +\psi_k j_\ell\l k\chi\r \nn\\&& + \int_0^{\chi}\d\tilde\chi\l\tilde\phi'_k+\tilde\psi'_k\r j_\ell\l k\tilde\chi\r\bigg]\Bigg\}
\,, \label{eq:angHI}
\eea
where $\delta^{\rm s}_k$ is the dark matter density contrast in the matter rest frame, $\cal H$ is the conformal Hubble parameter, $\chi$ is the comoving line-of-sight distance, and $v_k$ is the peculiar velocity, $\psi_k$ and $\phi_k$ are the metric potentials which for $\Lambda$CDM and standard dark energy models are equal, and $j_\ell$ are the spherical Bessel functions. Note that the first line corresponds to the density and RSD contributions common in the literature. The following lines include the so called GR corrections. We introduced the fudge factor $\epsilon_{GR}=1$ to identify the effects of such corrections. It is well know that they only affect very large scales since they have a $\propto k^{-2}$ dependence. The exception is the Doppler term $v_k j_\ell'$ which has a $\propto k^{-1}$ dependence and is the dominant GR correction in HI IM.

For the clustering bias $b^{\hi}$ we will follow \cite{Bacon:2018dui} and use the parametrisation
\be
b^{\rm \hi}(z)=0.667 + 0.178 z + 0.050 z^2\,.
\ee
In the case of HI IM the evolution bias can be written in terms of the temperature
\be 
b_e^{\rm \textsc{Hi}}(z)=-\frac{\p\ln \big[T^{\rm \textsc{Hi}}(z) H\z\big] }{\p\ln (1+z)}-2\,,
\ee
which we take to be expressed as \citep{Bacon:2018dui}
\be
\bar{T}^{\hi}(z)=0.056 + 0.232 z - 0.024 z^2\,.
\ee

In Eq. \ref{eq:angHI} the HI clustering bias receives a scale dependent correction $\Delta b^{\hi}$ if there is a non-zero primordial local-type non-Gaussianity. One can show that such correction is well approximated by \citep{Dalal:2007cu,Matarrese:2008nc}
\be\label{eq:Delta_bHI}
\Delta b^{\hi} =3\fnl \frac{\big[ b^{ \hi}(z) -1\big]\Omega_m H_0^2 \delta_c}{D(z) T(k) k^2},
\ee
where $\delta_c\simeq 1.69$ is the critical matter density contrast for spherical collapse, $T$ is the matter transfer function (normalised to 1 on large scales) and $D$ is the growth factor (normalised to 1 at $z=0$). This correction becomes important on very very large scales ($k\rightarrow0$) where $T(k)\simeq 1$ and therefore $\Delta b^{\hi}\propto \fnl k^{-2}$. We can then see that the effect of primordial non-Gaussianity as a similar scale dependence as the GR light-cone effects. In fact, neglecting such corrections may lead to spurious detections of primordial non-Gaussianity \citep{Camera:2014bwa}.

To compute the angular power spectrum we used a modified version of the publicly available code CAMB sources \citep{Challinor:2011bk} to include $\fnl$ \citep{Camera:2013kpa}. We plot three examples of the cosmological signal in cyan in Figure \ref{fig:obs_cl_foreg}.

\bfig
\centering
\includegraphics[width=\columnwidth]{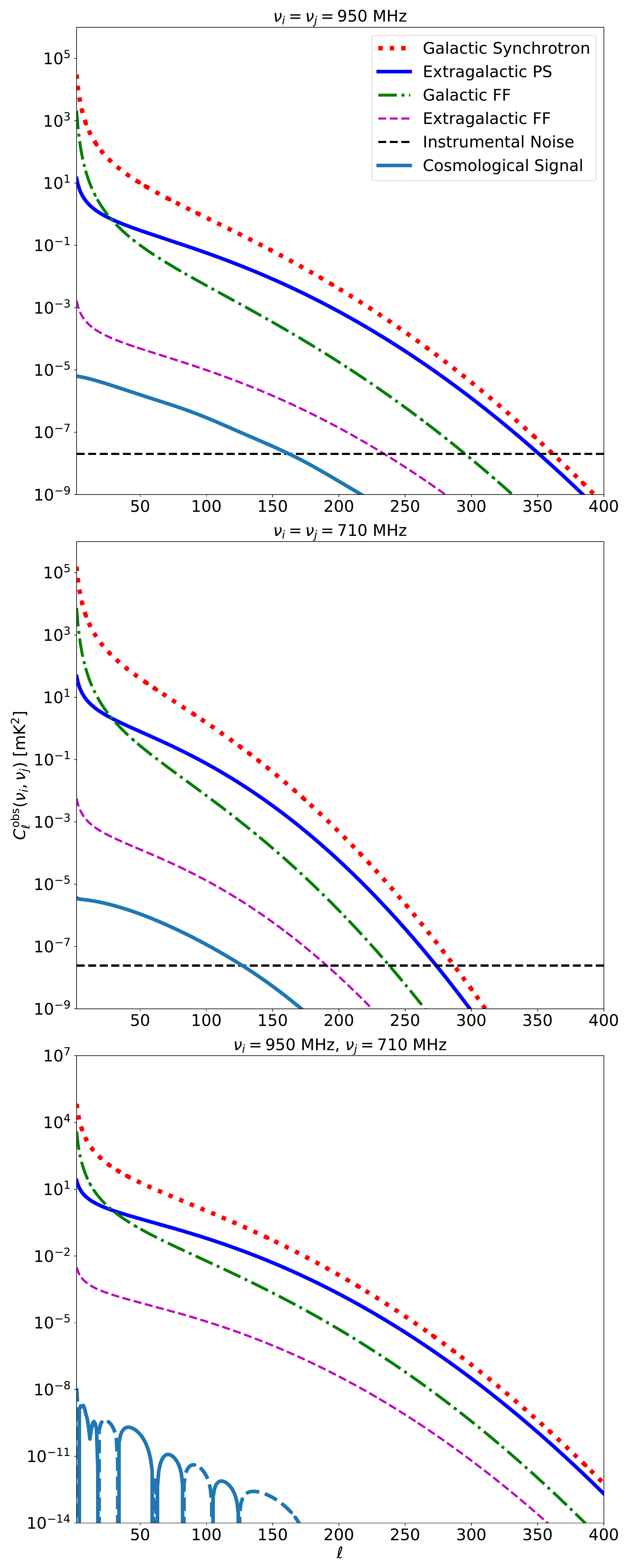}
\caption{Angular power spectrum of each component computed at $\nu_i=950$ MHz (\emph{Top}), at $\nu_j=710$ MHz (\emph{Middle}), and considering the cross correlation between the two different frequencies (\emph{Bottom}). Note that the cross-correlation between frequency bins has no instrumental noise contribution. The cross-correlation of the cosmological signal is plotted with a dashed line if its value is negative.}
\label{fig:obs_cl_foreg}
\efig

\subsection{Foregrounds}
\label{sec:foregrounds}

\begin{table}
\caption{\label{tab:for_pars} Calibrations of different foregrounds using $\nu_{\rm ref}=130$ MHz and $\ell_{\rm ref}=1000$ from \citet{Santos:2004ju}.}
\centering
\begin{tabular}{lcccc}
\hline
Foreground  & $\cal A$ [mK$^2$] & $\beta$ & $\alpha$  & $\xi$\\ 
\hline
Extragalactic Point Sources & 57.0 & 1.1 & 2.07  & 1.0\\
Extragalactic Free-Free & 0.014 & 1.0 & 2.10  & 35\\
Galactic Synchrotron & 700 & 2.4 & 2.80  & 4.0 \\
Galactic Free-Free & 0.088 & 3.0 & 2.15  & 35 \\
\hline
\end{tabular}
\end{table}

Our main goal is to assess how a joint fit to the cosmological parameters and the foregrounds would degrade our constraints. For that we need to establish models for the angular structure of the foregrounds. We follow \citet{Santos:2004ju} and use the generic expression
\be \label{eq:foregMS05}
C^{\rm MS05}_\ell (\nu_i,\nu_j)= {\cal A} \l\frac{\nu_{\rm ref}^2}{\nu_i\nu_j}\r^\alpha \  \l\frac{\ell_{\rm ref}}{\ell}\r^{\beta}\ e^{-\frac{\l \log \nu_i/\nu_j\r^2}{2\xi^2}} \,.
\ee
for the different foregrounds we consider. Here we will consider four different foreground components: Galactic free-free (GFF) emission, Galactic synchrotron (GS) emission, extragalactic free-free (EFF) and extragalactic point sources (EPS). In table \ref{tab:for_pars} we specify the fiducial calibrations for the references $\nu_{\rm ref}=130$ MHz and $\ell_{\rm ref}=1000$ from \citet{Santos:2004ju}. From the amplitudes of the power spectrum in Table \ref{tab:for_pars} one can clearly see that synchrotron is expected to be the dominant contributor. As for the signal, we need to convolve the foregrounds power spectra with the telescope beam, i.e.,
\be \label{eq:foregobs}
C^{\cal F}_{\ell,ij}=B_{\ell,i} B_{\ell,j} C^{\rm MS05}_{\ell, ij}\,.
\ee 
We can find examples of the angular power spectrum of the foregrounds in Figure \ref{fig:obs_cl_foreg} with Galactic Synchrotron in dotted red, extragalactic point sources in solid blue, Galactic free-free in dot-dashed green and extragalactic free-free in dashed magenta.

\subsection{Instrumental Noise}
\label{sec:noise}

Assuming uncorrelated Gaussian instrumental noise we have that 
\be \label{eq:inst_noise_hiim}
C_{\ell,ij}^{\cal N} =\frac{4\pi\,f_{\rm sky}\, T_{\rm sys,i}^2}{2\, N_{\rm d}\, \Delta \nu\, t_{\rm tot}}
\, \delta_{ij}\,.
\ee 
We will take the same system temperature specifications as in \citep{Bacon:2018dui} for both band 1 and band 2 of SKA1-MID. We will consider a 20 000 $\deg^2$ survey over 10 000 hours. We will round the number of dishes that constitute the observatory, taking $N_d=200$. We will consider fixed size frequency bins of 15 MHz, which correspond to a $\Delta z$ that ranges from 0.018 at low redshifts to 0.22 in the highest redshift bin. In figure \ref{fig:obs_cl_foreg} we show the instrumental noise in the horizontal dashed black line for two frequencies bin. Note that we assume that the noise power spectrum is uncorrelated in frequency, hence the bottom panel of \ref{fig:obs_cl_foreg} has no instrumental noise.

\bfig
\centering
\includegraphics[width=\columnwidth]{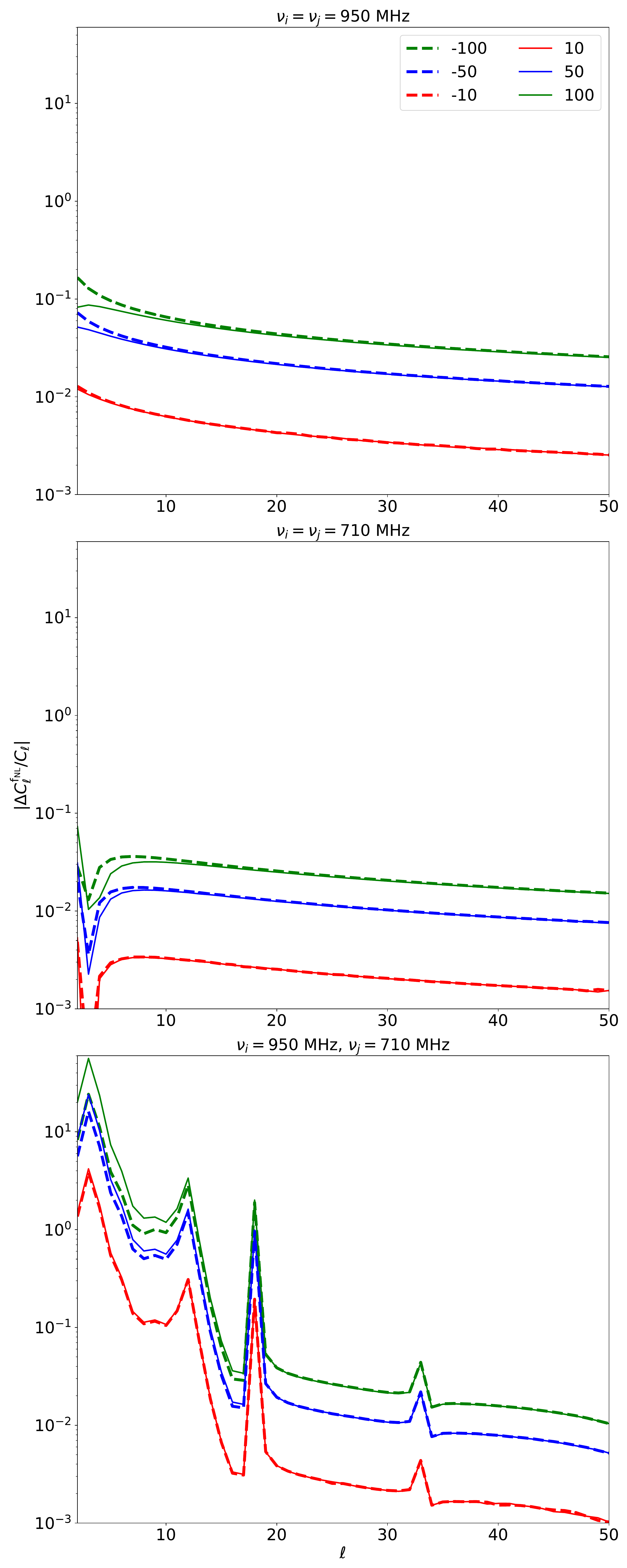}
\caption{Absolute value of the relative relevance of different PNG values of $\fnl=(-100,-50,-10,10,50,100)$ with respect to the Gaussian assumption for a 15MHz frequency bin at $\nu_i=950$ MHz (\emph{Top}), at $\nu_j=710$ MHz (\emph{Middle}), and considering the cross correlation between the two different frequencies (\emph{Bottom}). We define $\Delta C^{\fnl}_\ell\equiv C_\ell(\fnl) -C_\ell(\fnl=0)$.}
\label{fig:fnl_effect}
\efig

\bfig
\centering
\includegraphics[width=\columnwidth]{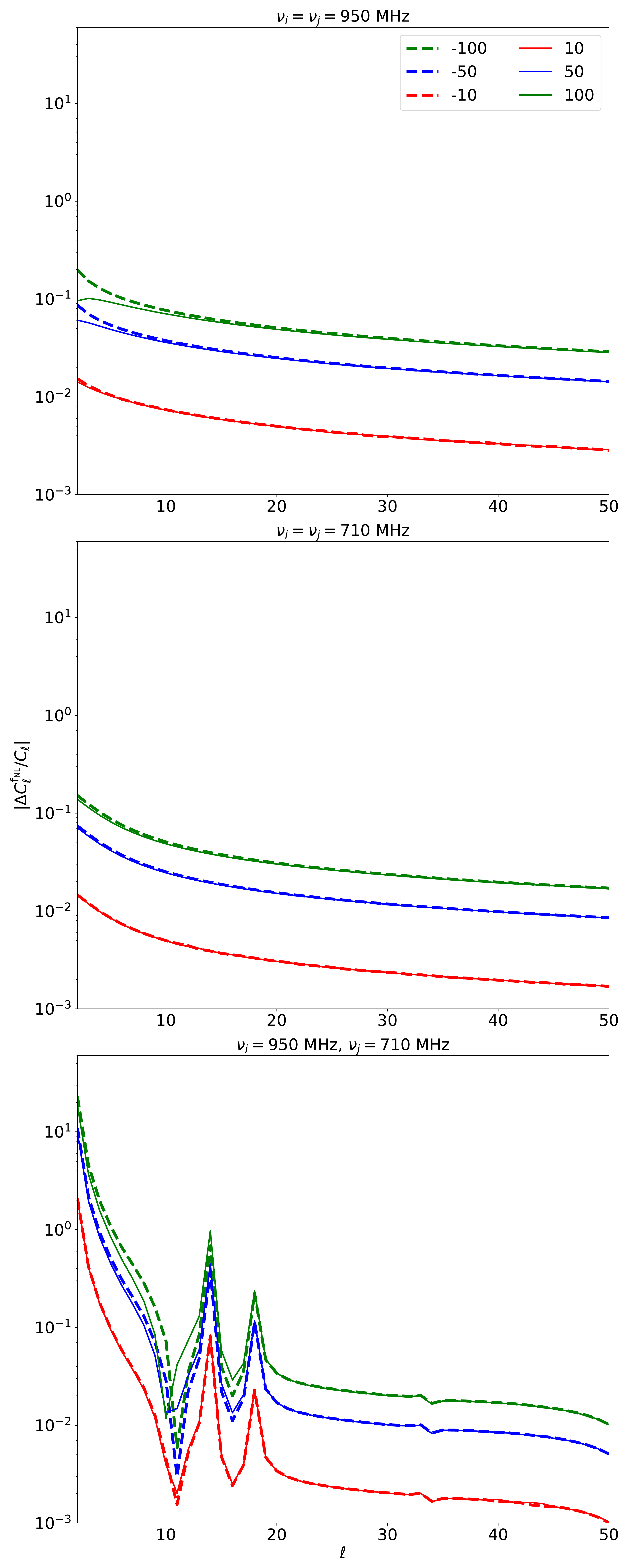}
\caption{Same as Figure \ref{fig:fnl_effect} but for frequency bins of 20MHz.}
\label{fig:fnl_effect_20MHz}
\efig

\subsection{The observed angular power spectrum}

To exemplify the importance of each contribution to the total observed angular power spectra is given by Eq. \ref{eq:cl_map}, we choose two reference frequencies $\nu_i=950$ and $\nu_j=710$ MHz ($z_i=0.49$ and $z_j=1.0$). In Figure \ref{fig:obs_cl_foreg} we plot the cosmological signal and the different foreground components convolved with the telescope beam for the two reference frequencies, as well as the instrumental noise. We also plot the cross-correlation angular spectra between the two frequencies. One can see the hierarchy in contributions to the total signal. Synchrotron emission is overwhelmingly higher than any other, although on large angular scales (small $\ell$) point-sources and galactic free-free are boosted. Fundamentally the cosmological signal is several orders of magnitude below the foregrounds but still above the noise level up to some $\ell$ depending on the observed frequency (as seen in the top and middle plots of Fig \ref{fig:obs_cl_foreg}). Despite the fact that traditional approaches use blind foreground cleaning methods, one can make the question if it would be possible to jointly measure the cosmological signal together with the foregrounds, given that we are in a low noise regime. Especially if we want to measure PNG and GR effects. In addition, the cross-correlations between bins give ample information to constrain the spectral indexes and are not affected by instrumental noise. Note as well that higher multipoles get their power dumped due to the beam of the experiment. 

Therefore, the question is how well we need to measure the \hi power spectra on large scale to detect $\fnl$ and the Doppler term (the main GR contributor). Firstly let us quantify how sensitive is the power spectrum to the large scale effects. In figure \ref{fig:fnl_effect} we plot relative contribution of PNG to the power spectrum in 15MHz bins for $\fnl=(-100,-50,-10,10,50,100)$ where $\Delta C^{\fnl}_\ell\equiv C_\ell(\fnl) -C_\ell(\fnl=0)$. While for $\fnl=\pm10$ the effect is below percent level, for a higher value of $\fnl$ the difference is enormous, especially in the cross-correlations. In the case of the cross-bin power spectra, the effect of PNG can be enormous, even for small values of $\fnl$, as the cross-bin correlations are more sensitive to the large scale effects. Still, the low multipoles are the ones where the relative difference is higher. Roughly speaking, one needs to measure the power spectrum at a sub-percent level to detect the effect of PNG in a single angular power. Besides, this can be relaxed since we have several angular power spectra from which to get cumulative constraints on $\fnl$. In figure \ref{fig:fnl_effect_20MHz} we do the same exercise but for thicker bins. One can see that the specific numbers change although the conclusion is similar, the effect is larger on low multipoles with the cross-correlations between frequency bins being the most sensitive.

\bfig
\centering
\includegraphics[width=\columnwidth]{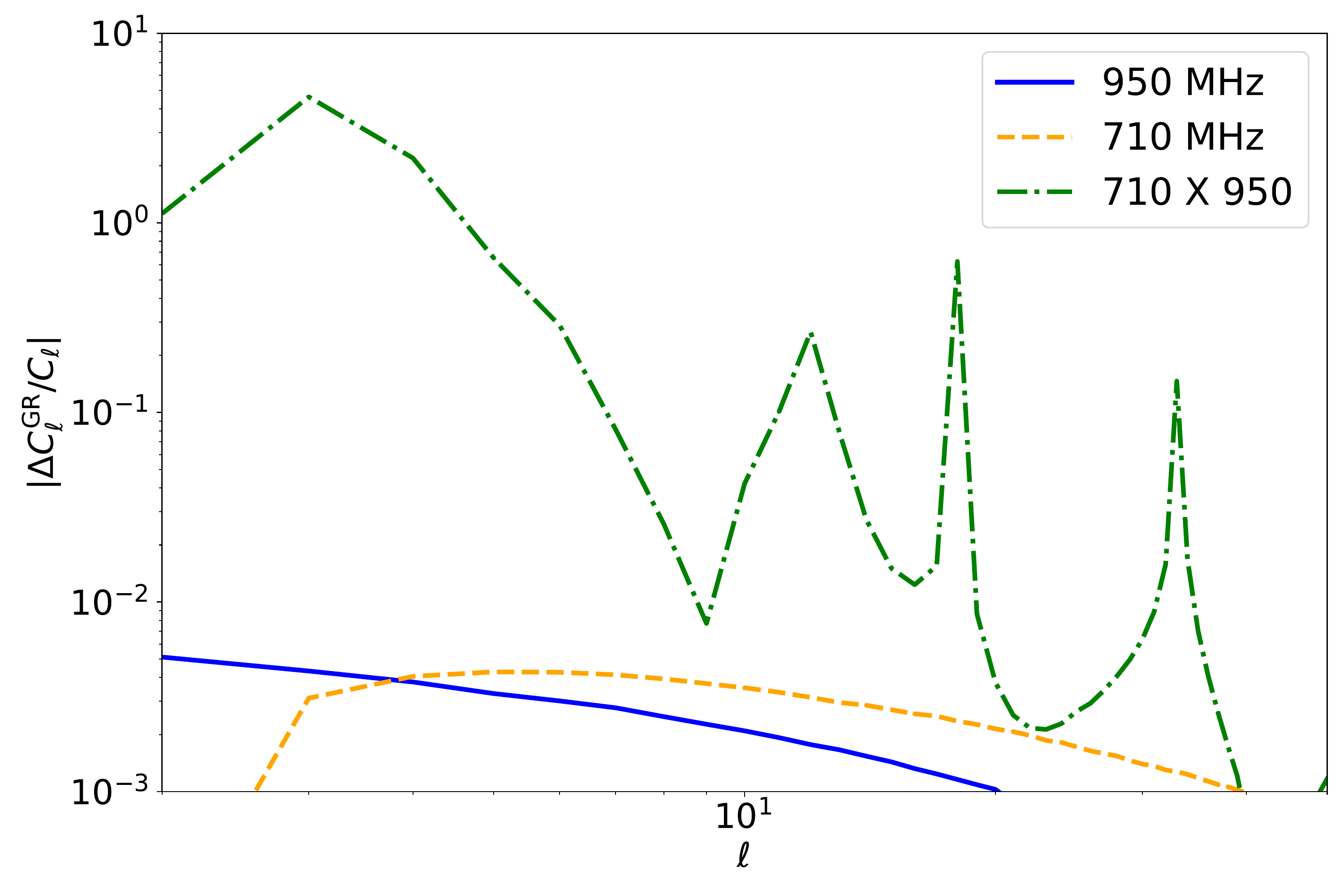}
\caption{Absolute value of the relative difference of the neglect of GR effects on the light-cone for frequency bins of 15MHz. We define the difference $\Delta C^{\rm GR}_\ell\equiv C_\ell(\epsilon_{\rm GR}=1) -C_\ell(\epsilon_{\rm GR}=0) $. }
\label{fig:gr_influence}
\efig

\bfig
\centering
\includegraphics[width=\columnwidth]{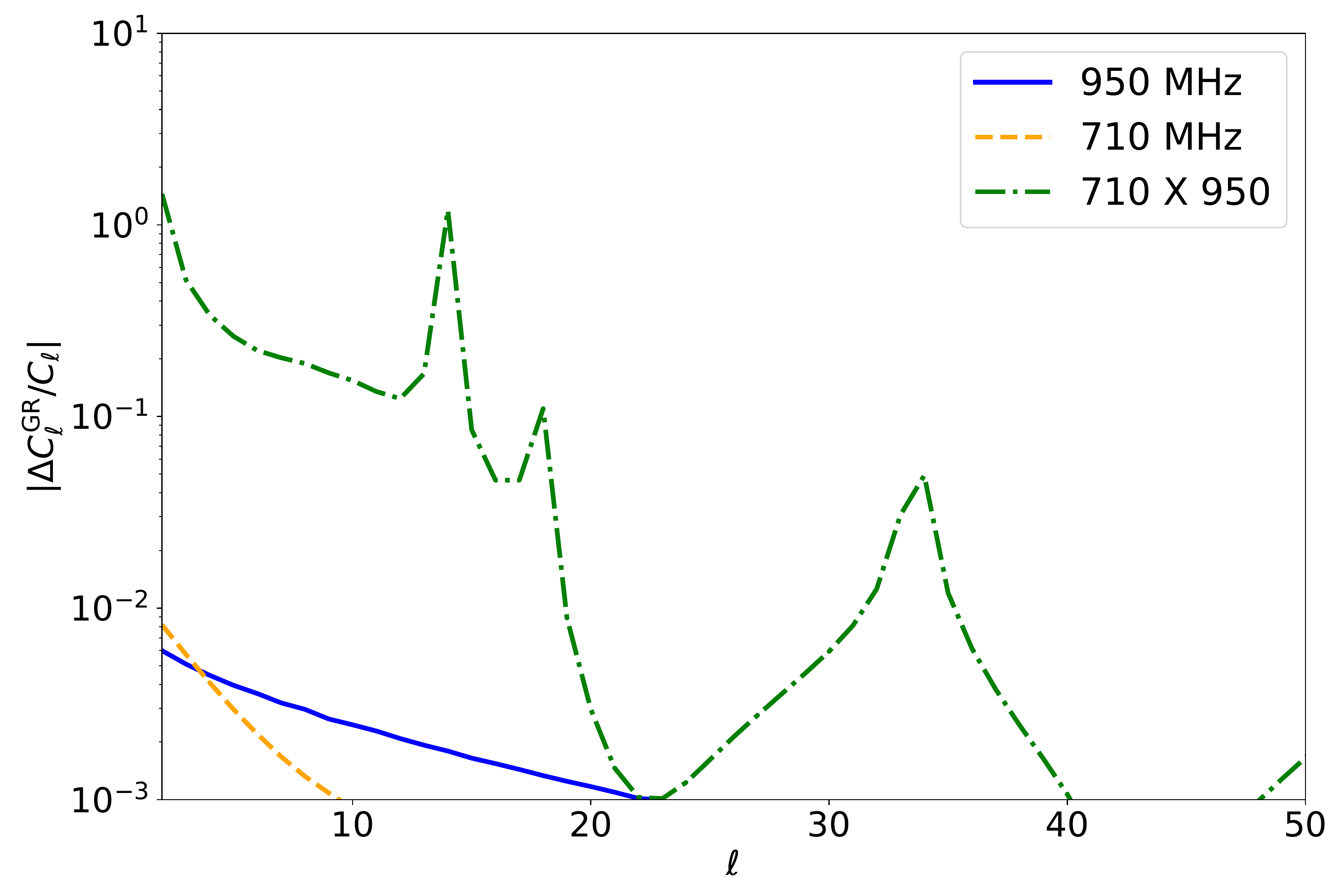}
\caption{Same as Figure \ref{fig:gr_influence} but for frequency bins of 20MHz. }
\label{fig:gr_influence20MHz}
\efig

In Figure \ref{fig:gr_influence} we plot the absolute value of the relative difference between including and neglecting the GR effects. While these effects are very small in the auto-bin correlations they become relevant in the cross-bin correlations, well above the percent level. Hence, one may hope that the cumulative signal of cross-bin correlations may allow us to identify the Doppler contribution. Besides, GR effects introduce $\ell$ structure in the cross-correlation which should be a distinctive signal and bigger than the effect of $\fnl$. Despite this the importance of the GR effects is dependent on the width of the bin, as we can see in Figure \ref{fig:gr_influence20MHz}. Larger frequency bins reduce the relevance of the so-called GR effects. This comes as no surprise, as the biggest of such effects is the Doppler term \citep{Challinor:2011bk} which is averaged out in thick redshift bins.

\section{The information matrix}
\label{sec:methods}

To estimate how well we can jointly measure the cosmological parameters and the foregrounds we will use the informations matrix which to leading order is given by \citep{Tegmark:1996bz}
\be \label{eq:finfo}
{\rm \bf F}_{\vartheta_\alpha\vartheta_\beta}=\sum_\ell \frac{\partial C^{\cal M}_{\ell,ij}}{\partial \vartheta_\alpha} \Gamma_{\ell,ij,mn}^{-1} \frac{\partial C^{\cal M}_{\ell,mn}}{\partial \vartheta_\beta}\,,
\ee
for a set of parameters  $\{\vartheta\}$. Then one determines the forecasted marginal error using 
\be
\sigma_\vartheta=\sqrt{\l {\rm \bf F}^{-1}\r_{\vartheta\vartheta}}\,
\ee
while the conditional error is 
\be
\sigma^{\rm cond}_\vartheta=1/\sqrt{{\rm \bf F}_{\vartheta\vartheta}}\,.
\ee
In the standard approach $C^{\cal M}$ is given by Eq. \ref{eq:cl_map} while $\Gamma_{\ell}$ is given by Eq. \ref{eq:cov}. In addition to the standard approach we want to compare it with two other scenarios. In the first scenario we want to understand how well we can learn physical and spatial properties of the foregrounds in the absence of cosmology. The goal is to set a base from which one can determine deviations from a foregrounds only scenario. In this \textit{no Cosmology} case, the covariance is simply given by the instrumental noise
\be 
\l\Gamma_{\ell,\ell'}\r_{ij,mn}^{\cal N} = \frac{\delta_{\ell,\ell'}}{(2\ell+1)\fsky} \l C^{{\cal N},im}_\ell C^{{\cal N},jn}_{\ell} +C^{{\cal N},in}_\ell C^{{\cal N},jm}_{\ell} \r \,.
\ee
Similarly the fiducial $C_\ell$ to input in Eq. \ref{eq:finfo} only has contributions from the noise and the foregrounds. By the same token, one can only forecast constraints on the foreground parameters but it will give us a ``best case scenario" for understanding the foregrounds and how degenerate are they among each other. 

In a second scenario, we will no longer consider that the foreground angular power spectrum has no covariance. The foregrounds are not a realisation of some foreground distribution function, they come from gas or charged particles from the galaxy or in the local Universe. Therefore they have no statistical structure. The point sources may be an exception. Still, let us assume that the parametrisation in Eq. \ref{eq:foregMS05} is incomplete or the model for the foregrounds has intrinsic small scale fluctuations. We thus model such uncertainty by introducing a foregrounds covariance linearly dependent on the foregrounds themselves, i.e. 
\be
\l\Gamma_{\ell,\ell'}\r_{ij,mn}^{\epsilon}= \frac{\delta_{\ell,\ell'}}{(2\ell+1)\fsky}\l C^{{ \epsilon},im}_\ell C^{\epsilon,jn}_{\ell} + C^{\epsilon,in}_\ell C^{{\epsilon},jm}_{\ell} \r
\ee
with
\be \label{eq:cov_eps}
C^{{ \epsilon},im}_\ell = C_\ell^{\cal S}\l \nu_i,\nu_j\r+\epsilon C_\ell^{\cal F}\l \nu_i,\nu_j\r+C^{\cal N}_\ell\l \nu_i,\nu_j\r \,,
\ee
Note that we will still consider each component to be uncorrelated with each other. In a nutshell, we made the covariance larger widening the error bars. One should also note that $\epsilon$ cannot be bigger than the uncertainty from the noise itself ($\epsilon\sim {\cal N}/{\cal A_F}$), as that sets the uncertainty in measuring the angular power spectrum of the foregrounds. Looking at Figure \ref{fig:obs_cl_foreg} this is highly dependent on the scale and may vary between $10^{-12}-10^{-6}$. In other words, our model uncertainties should not be bigger than the noise itself, otherwise one would change the foregrounds model. An alternative approach would be to insert stochastic nuisance parameters in every $\ell$ in every correlation. In practice, this is unfeasible by the sheer number of stochastic parameters one would need to introduce.  

In this paper we will consider both cosmological and foregrounds parameters. For the cosmological parameters we will consider the standard parameters, plus PNG and the GR fudge factor introduced in Eq. \ref{eq:angHI}, i.e.,  
\be\label{par_cosm}
\vartheta_{Cosmology} = \big\{ A_s, n_s,\Omega_{\rm CDM}, \Omega_{\rm b}, w, H_0, f_{\rm NL}, \epsilon_{\rm GR}  \big\}\nn\,,
\ee  
together with the biases is each bin $b(z_i)$ as nuisance parameters. We considered the following fiducial values for the cosmological parameters:  $A_s=2.142\times10^{-9}$, $n_s=0.9667$ $\Omega_{\rm CDM}=0.26$, $\Omega_{\rm b}=0.05$, $w=-1$, $H_0=67.74$km/s/Mpc. We also took $f_{\rm NL}=0$ (see Eq. \ref{eq:Delta_bHI}) and $\epsilon_{\rm GR}=1$ (see Eq. \ref{eq:angHI}). As foreground parameters we will consider the 4 free parameters in the model described by Eq. \ref{eq:foregMS05} for each of the 4 foreground components, i.e., 
\bea\label{par_forg}
\vartheta_{Foregrounds} &= \big\{& {\cal A}_{EPS}, \beta_{EPS}, \alpha_{EPS}, \xi_{EPS},\nn\\ 
&& {\cal A}_{EFF}, \beta_{EFF}, \alpha_{EFF}, \xi_{EFF}, \nn\\
&& {\cal A}_{GS}, \beta_{GS}, \alpha_{GS}, \xi_{GS}, \nn\\
&& {\cal A}_{GFF}, \beta_{GFF}, \alpha_{GFF}, \xi_{GFF}\quad \big\}\nn\,.
\eea 
The fiducials for the foregrounds are given in Table \ref{tab:for_pars}.

\section{Forecasting Results}
\label{sec:results}

In all the forecasts presented here we decided to truncate Eq. \ref{eq:finfo} at $\ell_{max}=150$. Although we include the beam, in principle one can consider higher multipoles. This does not add substantial information to cosmological constraints and increases the computation time. We will take $\ell_{\rm min}=3$ as the sky area allows it and we are considering the foregrounds in our fit. We also binned the full SKA1-MID into 15MHz slices, this way we ensure that in the lowest redshift bin is thick enough such that we are still in the linear regime along the line-of-sight, while trying to take advantage of the frequency resolution of the SKA1-MID. This gives 63 frequency channels with a total of 2016 independent angular power spectra.

\begin{table}
\caption{ Forecasted relative 1-$\sigma$ marginal errors ($\sigma_\vartheta/\vartheta$) in percentage ($\%$) for each foreground parameter under different assumptions. \label{tab:foreg_forecasts}}
\centering
\begin{tabular}{lcccc}
\hline
 & ${\cal A}_{GS}$ & $\beta_{GS}$ & $\alpha_{GS}$ & $\xi_{GS}$ \\ 
\hline
w/o cosmology & 7.3\e-9 &5.5\e-10 &5.4\e-10 &9.5\e-10\\
w/ cosmology & 2.1\e-05 & 1.7\e-08 & 1.5\e-08 & 3.3\e-08\\
$\log \epsilon=-12$ & 4.3\e-06 & 4.6\e-07 & 4.1\e-07 & 4.4\e-07\\
$\log \epsilon=-9$ & 4.5\e-06 & 4.8\e-07 & 4.3\e-07 & 4.5\e-07\\
$\log \epsilon=-6$ & 9.3\e-04 & 1.4\e-04 & 3.6\e-05 & 2.6\e-05\\
\hline
 & ${\cal A}_{GFF}$ & $\beta_{GFF}$ & $\alpha_{GFF}$ & $\xi_{GFF}$ \\ 
\hline
w/o cosmology & 6.6 \e-07 & 2.0\e-08 & 2.5\e-08 & 2.4\e-08\\
w/ cosmology & 2.2\e-05 & 6.4\e-07 & 9.0\e-07 & 8.8\e-07\\
$\log\epsilon=-12$ & 7.0\e-04 & 2.6\e-05 & 2.9\e-05 & 2.8\e-05\\
$\log\epsilon=-9$ & 7.2\e-04 & 2.7\e-05 & 3.0\e-05 & 2.8\e-05\\
$\log\epsilon=-6$ & 5.4\e-02 & 3.3\e-03 & 2.8\e-03 & 2.4\e-03\\
\hline
 & ${\cal A}_{EPS}$ & $\beta_{EPS}$ & $\alpha_{EPS}$ & $\xi_{EPS}$ \\ 
\hline
w/o cosmology & 8.2\e-05 &3.4\e-06 &9.3\e-07 &1.8\e-05\\
w/ cosmology & 4.0\e-03 & 1.7 \e-04 & 3.8\e-05 & 8.7\e-04 \\
$\log\epsilon=-12$ & 4.8\e-03 & 2.0\e-04 & 7.7\e-05 & 1.1\e-03\\
$\log\epsilon=-9$ & 5.6\e-03 & 2.3\e-04 & 8.6\e-05 & 1.3\e-03\\
$\log\epsilon=-6$ & 1.7\e-02 & 1.7\e-03 & 9.1\e-04 & 3.8\e-03\\
 \hline
 & ${\cal A}_{EFF}$ & $\beta_{EFF}$ & $\alpha_{EFF}$ & $\xi_{EFF}$ \\ 
\hline
w/o cosmology & 3.3\e-01 & 2.0\e-02 &5.7\e-03 &9.7\e-02\\
w/ cosmology & 1.6\e01 & 9.4\e-01 & 2.7\e-01 & 4.8\e-00\\
$\log\epsilon=-12$ & 1.9\e01 & 1.5\e00 & 4.6\e-01 & 5.7\e00\\
$\log\epsilon=-9$ & 2.2\e01 & 1.7\e00 & 5.1\e-01 & 6.7\e00\\
$\log\epsilon=-6$ & 1.4\e02 & 4.0\e01 & 1.1\e01 & 1.9\e01 \\
\hline
\end{tabular}
\end{table}

In table \ref{tab:foreg_forecasts} present our forecasts for the foreground parameters. We consider 5 different cases:
\begin{enumerate}
\item the first case we assume that the covariance of the estimator of the angular power spectrum has only instrumental noise. In this case we pretend to create a baseline for how well one we would constraint the foregrounds with our experiment and survey in the absence of cosmological signal. We called this case ``without" cosmology;\\
\item  the second case is the standard case where we include the \hi power spectrum in the covariance as computed in Appendix \ref{apx:cov_cl}. We therefore called this case ``with cosmology";\\
\item the further 3 cases include uncertainties in the modelling of the foregrounds by adding a contribution to the estimator covariance modulated by the parameter $\epsilon$ (see Eq. \ref{eq:cov_eps}).  We chose the value of $\epsilon$ to be roughly the ratio of the foregrounds and the instrumental noise power spectra in different $\ell$-scales. We chose this ratio as any uncertainties in the foregrounds above the noise would be detectable. In practice such contribution increase the uncertainty budget.  
\end{enumerate} 

Generically one can constrain very well the foregrounds in all scenarios one considers. The exception is extragalactic free-free emission. In the normal scenario, one can only constrain the amplitude of EFF with 16\% and its other parameters around percent level. In general, one can conclude that we will be able to learn the foregrounds with high accuracy even when we include foreground model uncertainties. This result may come as no surprise, we input well-defined models of the foregrounds in an extremely low noise experiment therefore we expect to measure them very well! This is true even when we assumed model uncertainties to be present. The results in table \ref{tab:foreg_forecasts} are not the crucial conclusion to take home, instead one should look for extensive and realistic modelling of the foregrounds, as they can be statistically tested with high precision. Not only they would complement the information about dust and Synchrotron emission from higher frequencies \citep{Akrami:2018mcd} but also improve the constraints available. While for Planck there were 9 frequency channels available, the SKA1-MID is only limited by RFI flagging. Here we considered 63 which is almost an order of magnitude increase in the band available.     

\begin{table}
\caption{\label{tab:results_cosm} Forecasted relative 1-$\sigma$ marginal errors ($\sigma_\vartheta/\vartheta$) in percentage ($\%$) for the standard cosmological parameters under different assumptions.}
\centering
\begin{tabular}{lcccccc}
\hline
 & $A_s$ & $n_s$ & $H_0$ & $\Omega_{CDM}$ & $\Omega_{b}$ & w \\ 
\hline
w/o foregrounds & 3.57 & 1.42 & 2.31 & 2.32 & 4.05 & 2.41 \\
w/ foregrounds & 3.60 & 1.43 & 2.32 & 2.38 & 4.11 & 2.52 \\
$\log \epsilon=-12$ & 3.61 & 1.43 & 2.33 & 2.38 & 4.11 & 2.53 \\
$\log \epsilon=-9$ & 3.64 & 1.44 & 2.34 & 2.41 & 4.14 & 2.57 \\
$\log \epsilon=-6$ & 3.73 & 1.47 & 2.40 & 2.48 & 4.26 & 2.66 \\
\hline
\end{tabular}
\end{table}

The results for the standard cosmological parameters are presented in table \ref{tab:results_cosm}. As for table \ref{tab:foreg_forecasts}, the different lines correspond to the different cases we want to investigate (although we now do not consider the case with noise only). The most important conclusion to take is that the marginal constraints on the standard cosmological parameters are fairly independent of the foregrounds even when we increase the covariance. This is in agreement with previous results \citep[see for example][]{Wolz:2013wna}, as the foreground cleaning methods do not affect smaller scales. Note as well that we considered the bias as nuisance parameters but we assumed the \hi temperature to be known which can degrade considerably the constraints on the cosmological parameters. In principle one can assume that other summary statistic like the 3D power spectrum, or its multipoles, with foreground cleaning are used for constraining the standard cosmological model, which then one can take as ``known" when studying large scale effects.

Our main point for this paper was to understand how the foregrounds would affect $\fnl$. Let's first start with recapping the forecasts without foregrounds. The first three lines of table \ref{tab:results_fnl_gr} present the forecasts of the conditional error, the marginal error when we neglect the presence of foregrounds, and the case when we fix the bias. 
The marginal only degrades by a quarter and is not very sensitive to the knowledge of the bias. Note that the results presented here are different (and worse) from the results of \citet{Alonso:2015uua}. Here we include the beam and have a more stringent cut on the $\ell_{\rm max}$ (150 instead of 500), as well as using a smaller area and having more bias nuisance parameters. More than comparing forecasts we are interested in assessing how marginalising over the foregrounds degrades constraints. While one would hint that such an approach would render impossible any $\fnl$ measurement, this is not the case. Although the inclusion of the foregrounds degrades the constraints by $\sim75\%$, this is not catastrophic. In particular, one can devise a strategy where we use other methods to determine the cosmological parameters and only use the approach presented in this paper for the large scale effects and foreground parameterisations. Such a strategy would produce a marginally better $\sigma_{\fnl}=6.6$, which although far from the target $\lesssim1$, is comparable with other LSS experiments. Once we start making the covariance bigger due to foreground model uncertainties the forecasted error worsens. Similar conclusions can be taken for the GR effects. Still, it one clearly concludes that a joint fit is not unrealistic. 

\begin{table}
\caption{\label{tab:results_fnl_gr} Forecasted relative 1-$\sigma$ marginal errors ($\sigma_\vartheta$) for the large scale effects under different assumptions.}
\centering
\begin{tabular}{lcc}
\hline
 & $f_{NL}$ & $\epsilon_{GR}$ \\ 
\hline
w/o foregrounds & 4.81& 5.17\\
\quad+ bias fixed & 4.62 & 5.15 \\
w/ foregrounds & 8.50 & 6.70 \\
\quad+ bias fixed& 7.94 & 6.64\\
\quad+ cosmology fixed & 6.6 & 6.3\\
$\log \epsilon=-12$ & 8.70 & 6.79\\
$\log \epsilon=-9$ & 9.44 & 6.98\\
$\log \epsilon=-6$ & 10.84 &7.41\\
\hline
\end{tabular}
\end{table}

\section{Using the wrong models}
\label{sec:bias_results}

The conclusions we arrived at here can be reached just by looking at figure \ref{fig:obs_cl_foreg} without any calculation. If one neglects a contribution somewhere, it necessarily needs to be absorbed by other free parameters as it disappeared from the signal, irrespective of the volume of available data. Although all foregrounds included in our toy model were known, the results hint that any miss-modelling would be noticed straight away. In addition, the inclusion of fudge foreground component can be a test of unmodelled components. Hence a relevant question, which we have not considered yet, is whether a poor or incomplete modelling of the foreground components leads to significant biases in the final results. Considering how large is the foreground contribution, compared to the signal, this is a potentially crucial issue to investigate. In some sense including extra contributions to the covariance of the $C_\ell$ was an attempt to include issues with bad modelling, but they only degrade the precision of the measurement. This is quite clear in the results of the previous section. Irrespective of the precision, how is the accuracy of our measurement affected by using wrong models? 

To answer the question of how well we need to model the foregrounds, let us consider the case of nested model selection \citep{Heavens:2007ka}. Let us say that a bigger model $M$ has $\psi_i$ parameters and that the nested model $M'$ has $\theta_j$, meaning that $\phi_k$ remaining parameters are in the bigger model but not in the smaller. This means that $M'\subset M$ and $\{\psi_i\}=\{\theta_j\}\cup\{\phi_k\}$. If we fix the $\phi_k$ parameters at their correct values then the peak of the likelihood of the subset remains unchanged (although the shape of the posterior changes). On the other hand, if we fix them at an incorrect value $\Delta \phi_k$ away from the correct one, one can show (see Appendix \ref{apx:bias_par}) that we then bias the best fit value of the nested sample by
\be
\Delta\theta_j=-\left({\bf H}^{-1}\right)_{\theta_j\theta_i} {\bf F}_{\theta_i\phi_k}\Delta \phi_k \,,
\ee
where {\bf H} is is the fisher matrix of the simpler model while {\bf F} is the fisher matrix of the bigger model. Note that this expression is only valid for perturbations around the best fit model. In this paper, we will grossly extrapolate its validity to gain insight into how badly one can bias the best fit values. 

The first error one can make is to assume an incorrect number of foregrounds in the forward model. As expected such information would spill into the other components. Let us take for example that we have neglected extragalactic free-free, as it is the smallest of the foregrounds component. That means $\Delta \phi_k=\Delta {\cal A}_{EFF}=-0.014$. If we had done so, we would have biased our best fit parameter of the GR corrections and $\fnl$ by
\be
\Delta \fnl= 9687\,, \quad \Delta \epsilon_{GR}= 13460\,.
\ee
These numbers are slightly non-sensical in the sense that if we had measured them in reality one would have immediately understood that something had gone wrong. Still, the amplitude of a neglected component needs to be absorbed by any other free parameter. In this test case, not only the large scale effects are biased but all other cosmological parameters. Although the foreground parameters are biased they are minimally so. 

Another possible ``error" would be to assume that we only have galactic foregrounds. If this was the cases the biasing of the super horizon scale effects would be even more severe
\be
\Delta \fnl \sim \Delta \epsilon_{GR} \sim -10^9\,.
\ee
Note that the exact number is not needed as our approach has already broken down. But it does mean that missing foreground components will bias any fit of the cosmological parameters. In this scenario it is the galactic foregrounds that change substantially, even absorbing the cosmological signal,
\be
\Delta {\cal A}_{GS}= 12.4\,, \quad \Delta {\cal A}_{GFF} = 0.11 \,.
\ee
This would make the angular power of galactic free-free to more than double in amplitude. 

One could also assume that the foregrounds are either perfectly correlated or uncorrelated in frequency. If they are perfectly correlated, $\xi\rightarrow \infty$. In practice one cannot do this but we can make $\xi$ be large. One can estimate the value of $\xi$ that makes them correlated at least of in 1 in a thousand within the frequency range of the experiment $[350,1420]$ MHz. That requires a value of $\xi=31.3$ but let's make it equal to 35 for simplicity. Then $\Delta \phi_k=(35-\xi^{fid}_{EPS},35-\xi^{fid}_{EFF},35-\xi^{fid}_{GS},35-\xi^{fid}_{GFF})=(34,0,31,0)$. The results would be catastrophic for $\fnl$ and $\egr$, as well as for the fits to the foregrounds themselves. On the other hand, if we consider the foregrounds uncorrelated in frequency then $\xi\rightarrow 0$. In practice one cannot do this but one can take a small value of $\xi$, like $\xi=1$. Then $\Delta \phi_k=1-(\xi^{fid}_{EPS},\xi^{fid}_{EFF},\xi^{fid}_{GS},\xi^{fid}_{GFF})=-(0,34,3,34)$ would bias the best fit parameters of all others parameters.

\section{Discussion}
\label{sec:conclusions}

The main goal of this paper was to estimate how the presence of foregrounds degrades forecasted constraints on $\fnl$ and GR effects using \hi IM. In this paper, we took an alternative approach to what is commonly found in the literature. Instead of using foreground cleaning methods, which highly suppresses information on very large scales, we use toy models of foregrounds to estimate which results can be obtained from a joint fit of foreground parameterisations, cosmological parameters, and large scale effects. We, therefore, started by reviewing maps of intensity, and the summary statistics we use in this paper, the angular power spectra $C_\ell$. We reviewed the theoretical covariance of the power spectra and argued that it should be independent of the foregrounds. We then modelled each component: the cosmological signal of HI, the foregrounds, and the instrumental noise. We also exemplified how the foregrounds and the cosmological signal compare with each other. Despite the wide amplitude differences between each contribution, i.e., the cosmological signal being subdominant with respect to any of the foreground components, the instrumental noise (plus the cosmology) sets the uncertainty one can measure the total angular power. We then exemplified the effect of large scale effects in the power spectrum to gain insight on how well one needs to measure the cosmological contribution in the total observed power spectra. 

We then reviewed the fisher matrix formalism and set up the parameters of interest as well as their fiducial values. We considered 3 ``experimental setups": one in which only the noise is relevant in the covariance, the traditional one where the variance of the angular power is given by the instrumental noise and cosmic variance, and a third where we model Gaussian fluctuations of the foregrounds proportional to its amplitude. Each case was considered in our SKA1-MID experimental setup. One of the main conclusions is that irrespective of the case one should be able to learn very well the angular structure and tomographic structure of the foregrounds in frequency. This is fundamentally due to the low experimental noise and a high number of correlations possible. Although including foreground model uncertainties degrades the constraints on the model parameters, these are limited. Fundamentally such variations should be detectable in such low noise experiments. Also, being able to understand so well the foregrounds does not affect the forecasted constraints on the standard cosmological parameters. 

On the other hand, the large scale effects get degraded substantially in the presence of foregrounds. Although the forecasted error degrades around 75\% in the case of $\fnl$, the constraints on GR effects is only degraded by 30\%. Even though the foregrounds degrade the constraints, these are neither catastrophic neither represent a substantial source of degeneracy. If we compare the conditional error with a marginal error when the cosmology is known, the foregrounds only represent a 50\% degradation. In any case, the survey specs we used of 20000$\deg^2$ can provide Planck-level constraints even if marginalised over the foregrounds - although strong priors on the cosmological parameters are required. In the absence of foregrounds our forecasted constraints on $\fnl$ are of same order of magnitude of what was previously found by \citet{Alonso:2015uua}, still far from a desirable constraining power of $\sigma_\fnl\lesssim1$. We therefore require more futuristic experiments, cross-correlate with optical galaxy surveys \citep{Fonseca:2015laa} or add bispectra information \citep{Karagiannis:2018jdt}.

Hence, it seems potentially feasible to jointly fit the cosmology, foreground parameters, and primordial non-Gaussianity. Such ``feasibility" can be dismissive. We used toy models that are a strong theoretical prior. We, therefore, tried to gain insight into how biased one would be if our understanding of the foregrounds is wrong. We used nested models to quantify the biases. We, therefore, imagined what would happen if we neglected one or several components to the power spectrum. We concluded that it would be fundamentally catastrophic, not only for the large scale effects but also to what one can learn about the foregrounds. As one expects, if we do not model a contribution it needs to be absorbed by other components. A similar conclusion is obtained when we assume wrong correlations in frequency. What this means is that any unaccounted contribution to the power spectra will jeopardise our results.

This work seems to disagree with \citet{Liu:2011aa} on the possibility of constraining foreground properties. We argue that there is a substantial amount we can learn about the foregrounds' macroscopical properties. There are differences between this work and their work. Firstly we parametrise the summary statistics of the angular and frequency structure of foregrounds while \citet{Liu:2011aa} works in pixel space using physical descriptions of foreground emission. Secondly, we focused on the SKA1-MID at higher frequencies in a wider range. This means that we assume we have more data available at a lower instrumental noise. Our work was not intended to replicate theirs and a proper comparison would need to be done with their parameterisations and the instrumental and survey specifications of SKA1-LOW. We leave this to future work. 

What we concluded here is that one can indeed jointly measure foreground parameters and large scale effects in HI IM. Although the degeneracies between foregrounds and $\fnl$ degrade the constraining power this is not catastrophic. But one needs to carefully include the proper foregrounds model and components. This means that although we show that we can measure foregrounds and large-scale effects in their presence we should take this result with a slight pinch of salt. Firstly we considered toy models for the foregrounds and neglected systematics like 1/f noise, polarisation leakage, and beam asymmetries. We also considered well defined spectral indices for each foreground instead of taking their power spectrum. Neither we consider a proper map of the foregrounds with mask cuts. As we showed here, any incorrect modelling will jeopardise any attempt not only of understanding the foregrounds but more interestingly primordial non-Gaussianity. Still, the results here indicate that, as in CMB studies \citep{Aghanim:2019ame,2019arXiv191000483E}, it is worth resorting to template fitting for specific applications which require accurate reconstructions of the large angular scales in the survey. This will allows us to learn properties of the foregrounds and constrain $\fnl$. 

As a summary, we explored the potential of intensity mapping surveys to constrain primordial local-type non-Gaussianity. We find that, with realistic settings for forthcoming experiments the expected constraints are at the level of those already achievable by Planck. This can of course be improved with more futuristic settings. However, the main message of our work is that the constraints are not significantly degraded by foreground contamination, provided the templates are accurate enough. Therefore, a template fitting approach is a worth pursuing methodology on the large scales required for $\fnl$ and GR effects studies.

\section*{Acknowledgements}
We thank S. Camera for useful discussions on the parameters bias formalism. We thank Mel Irfan for useful discussion on foregrounds and their nature. We also thank Filippo Oppizzi, and M\'ario G. Santos for useful discussions. JF and ML were supported by the University of Padova under the STARS Grants programme {\em CoGITO: Cosmology beyond Gaussianity, Inference, Theory, and Observations}. JF was also supported by the UK Science \& Technology Facilities Council (STFC) Consolidated Grant ST/P000592/1. This work made use of the South African Centre for High-Performance Computing, under the project \emph{Cosmology with Radio Telescopes}, ASTRO-0945.

\section*{Data Availability}
Data and codes are available on request. The version of CAMB used is already available online \url{https://github.com/ZeFon/CAMB_sources_MT_ZF}.





\appendix

\section{The covariance of the angular power spectrum in full sky}\label{apx:cov_cl}

Here we review the calculation of the covariance of the observed angular power spectrum. As in Eq. \ref{eq:cl_estim} we consider the full sky estimator or the angular power spectrum to be
\be
\hat C_{\ell,ij}^{\cal M}=\frac1{2\ell+1}\sum_{m=-\ell}^{m=\ell} \frac12 \Big[\almi{i}^{\cal M}\almi{j}^{\cal M,*}+\almi{j}^{\cal M}\almi{i}^{\cal M,*}\Big]\,.
\ee
From Eqs. \ref{eq:map} and \ref{eq:alm} it follows that
\be
\almi{i}^{\cal M}=\almi{i}^{\cal S}+\almi{i}^{\cal F}+\almi{i}^{\cal N}\,,
\ee
and assuming the components are independent, i.e.,
\be \label{eq:prop_alm}
\langle \almi{i}^{\cal A} a_{\ell' m', j}^{\cal B,*}\rangle=\delta^{\cal AB} \delta_{\ell\ell'} \delta_{mm'} C^{\cal A}_{\ell,ij}\,.
\ee
Then the expected value of our estimator is given by 
\be \label{eq:expect_cl_map}
\langle \hat C_{\ell,ij}^{\cal M} \rangle =C_{\ell,ij}^{\cal M}=C_{\ell,ij}^{\cal S}+C_{\ell,ij}^{\cal F}+C^{\cal N}_{\ell,ij} \,,
\ee
which can be seen as a biased estimator of the signal with the bias being the foregrounds. Note that, since the different components are independent, the total angular power spectrum is just the sum of the angular power spectrum of each component.  

The covariance of our angular power spectrum estimator is defined as
\bea
Cov \left[\hat C^{\cal M}_{\ell,ij}, \hat C^{\cal M}_{\ell',pq}\right] &=&\l\Gamma_{\ell,\ell'}\r_{ij,pq}\,,\\
& \equiv& \langle \hat C^{\cal M}_{\ell,ij} \hat C^{\cal M}_{\ell',pq} \rangle - \langle \hat C^{\cal M}_{\ell,ij} \rangle \langle \hat C^{\cal M}_{\ell',pq} \rangle\nn\,.
\eea
The expected value of the angular power spectrum has already been given by Eq. \ref{eq:expect_cl_map}. The first term becomes 
\bea \label{eq:2pCl}
\langle \hat C^{\cal M}_{\ell,ij} \hat C^{\cal M}_{\ell',pq} \rangle=&&\frac1{(2\ell+1)(2\ell'+1)}\sum_{m=-\ell}^{m=\ell}\sum_{m'=-\ell'}^{m'=\ell'} \frac14 \times \nn \\
&\Bigg\langle&\Big[\almi{i}^{\cal M}\almi{j}^{\cal M,*}+\almi{j}^{\cal M}\almi{i}^{\cal M,*}\Big]\times\nn\\
&&\Big[\almpi{p}^{\cal M}\almpi{q}^{\cal M,*}+\almpi{q}^{\cal M}\almpi{p}^{\cal M,*}\Big]\Bigg\rangle\,.\nn\\
&&
\eea
It is a summation over 4 similar 4-point functions. For simplicity let us just analyse the first one
\be
\left\langle \almi{i}^{\cal M}\almi{j}^{\cal M,*} \almpi{p}^{\cal M}\almpi{q}^{\cal M,*}\right\rangle\,. \nn
\ee
One would be tempted to use Wick's theorem immediately and just expand this 4-point function. But Wick's theorem in its simplest form is only valid for quantities that follow some distribution function. The noise is usually taken to be Gaussian, i.e., follows as Gaussian distribution with zero mean and covariance given by $C_\ell^{\cal N}$. The same is true for cosmological perturbations, it is a Gaussian random variable with covariance $C_\ell^{\cal S}$. On the other hand, the foregrounds are not a realisation of any intrinsic spatial distribution function. They are simply a local offset and $C_\ell^{\cal F}$ is just a characterisation of the spacial structure of the foregrounds in spherical harmonic space. 
Technically speaking, $\alm^{\cal F}$ gets out of the expected value brackets. Noting that Gaussian distributions have zero value odd-point functions we write
\bea\label{eq:2p_gauss}
\left\langle \almi{i}^{\cal M}\almi{j}^{\cal M,*} \almpi{p}^{\cal M}\almpi{q}^{\cal M,*}\right\rangle&=& \left\langle \almi{i}^{\cal G}\almi{j}^{\cal G,*} \almpi{p}^{\cal G}\almpi{q}^{\cal G,*}\right\rangle\nn\\
&+&\almi{i}^{\cal F}\almi{j}^{\cal F,*} \left\langle \almpi{p}^{\cal G}\almpi{q}^{\cal G,*}\right\rangle\nn\\
&+&\left\langle \almi{i}^{\cal G}\almi{j}^{\cal G,*} \right\rangle\almpi{p}^{\cal F}\almpi{q}^{\cal F,*}\nn\\
&+& \almi{i}^{\cal F}\almi{j}^{\cal F,*} \almpi{p}^{\cal F}\almpi{q}^{\cal F,*}\nn\\
\eea
where we defined the Gaussian part as $\almi{i}^{\cal G}\equiv \almi{i}^{\cal S} + \almi{i}^{\cal N}$. Hence when we do the summation in $m$ and $m'$ in Eq. \ref{eq:2pCl} we get for the last 3 terms of Eq. \ref{eq:2p_gauss}
\bea
\sum_{m=-\ell}^{m=\ell}\sum_{m'=-\ell'}^{m'=\ell'} \frac{\almi{i}^{\cal F}\almi{j}^{\cal F,*}}{(2\ell+1)} \frac{\left\langle \almpi{p}^{\cal G}\almpi{q}^{\cal G,*}\right\rangle}{(2\ell'+1)}   &=& C^{\cal F}_{\ell,ij} C^{\cal G}_{\ell',pq} \nn\\
\sum_{m=-\ell}^{m=\ell}\sum_{m'=-\ell'}^{m'=\ell'} \frac{\left\langle\almi{i}^{\cal G}\almi{j}^{\cal G,*}\right\rangle}{(2\ell+1)} \frac{ \almpi{p}^{\cal F}\almpi{q}^{\cal F,*}}{(2\ell'+1)}   &=& C^{\cal G}_{\ell,ij} C^{\cal F}_{\ell',pq} \nn\\
\sum_{m=-\ell}^{m=\ell}\sum_{m'=-\ell'}^{m'=\ell'} \frac{\almi{i}^{\cal F}\almi{j}^{\cal F,*}}{(2\ell+1)} \frac{\almpi{p}^{\cal F}\almpi{q}^{\cal F,*}}{(2\ell'+1)}   &=& C^{\cal F}_{\ell,ij} C^{\cal F}_{\ell',pq}\nn \\
&&
\eea
The factors of 1/4 vanish as well since we have 4 combinations of the $\alm$ in Eq. \ref{eq:2pCl}. We can now expand the Gaussian part of the 4 point function using Wick's theorem as
\bea
\left\langle \almi{i}^{\cal G}\almi{j}^{\cal G,*} \almpi{p}^{\cal G}\almpi{q}^{\cal G,*}\right\rangle&=& \left\langle \almi{i}^{\cal G}\almi{j}^{\cal G,*} \right\rangle \left\langle \almpi{p}^{\cal G}\almpi{q}^{\cal G,*}\right\rangle\nn\\
&+& \left\langle \almi{i}^{\cal G}  \almpi{q}^{\cal G,*} \right\rangle \left\langle \almpi{p}^{\cal G}  \almi{j}^{\cal G,*}\right\rangle\nn\\
&+& \left\langle \almi{i}^{\cal G} \almpi{p}^{\cal G} \right\rangle \left\langle\almpi{q}^{\cal G,*} \almi{j}^{\cal G,*}\right\rangle\,.\nn\\
\eea
Using Eq. \ref{eq:prop_alm} and the property $\alm^*=a_{\ell,-m}$ we get 
\bea
\left\langle \almi{i}^{\cal G}\almi{j}^{\cal G,*} \almpi{p}^{\cal G}\almpi{q}^{\cal G,*}\right\rangle&= &C^{\cal G}_{\ell,ij}C^{\cal G}_{\ell',pq}\nn\\
&+&C^{\cal G}_{\ell,iq}C^{\cal G}_{\ell,pj} \delta_{\ell\ell'} \delta_{mm'}\nn\\
&+&C^{\cal G}_{\ell,ip}C^{\cal G}_{\ell,qj} \delta_{\ell\ell'} \delta_{m,-m'}\,.\nn\\
\eea
We can then make the summation in $m$ and $m'$
\bea
&\sum_{m,m'} \frac{\left\langle \almi{i}^{\cal G}\almi{j}^{\cal G,*} \almpi{p}^{\cal G}\almpi{q}^{\cal G,*}\right\rangle}{(2\ell+1)(2\ell'+1)} = C^{\cal G}_{\ell,ij}C^{\cal G}_{\ell',pq}\nn\\
& + \frac{\delta_{\ell\ell'}}{(2\ell+1)}\left[C^{\cal G}_{\ell,iq}C^{\cal G}_{\ell,pj} + C^{\cal G}_{\ell,ip}C^{\cal G}_{\ell,qj} \right]\,.
\eea
Therefore the 2 point function of the angular power spectrum is
\bea
\left\langle \hat C^{\cal M}_{\ell,ij} \hat C^{\cal M}_{\ell',pq}\right \rangle &=& C^{\cal G}_{\ell,ij}C^{\cal G}_{\ell',pq} + C^{\cal F}_{\ell,ij} C^{\cal F}_{\ell',pq} \nn\\
&+&C^{\cal G}_{\ell,ij} C^{\cal F}_{\ell',pq} + C^{\cal F}_{\ell,ij} C^{\cal G}_{\ell',pq} \nn\\
&+& \frac{\delta_{\ell\ell'}}{(2\ell+1)}\left[C^{\cal G}_{\ell,iq}C^{\cal G}_{\ell,pj} + C^{\cal G}_{\ell,ip}C^{\cal G}_{\ell,qj} \right]\,.\nn\\
\eea
The first two lines are just $\langle \hat C^{\cal M}_{\ell,ij} \rangle \langle \hat C^{\cal M}_{\ell',pq} \rangle$, then we arrive to the covariance of our estimator to be 
\be
\l\Gamma_{\ell,\ell'}\r_{ij,pq} = \frac{\delta_{\ell\ell'}}{(2\ell+1)}\left[C^{\cal G}_{\ell,iq}C^{\cal G}_{\ell,pj} + C^{\cal G}_{\ell,ip}C^{\cal G}_{\ell,qj} \right]\,,
\ee
where $C_{\ell,ij}^{\cal G}=C_{\ell,ij}^{\cal S}+C^{\cal N}_{\ell,ij}$. Therefore the covariance only depends on the angular power spectra of the cosmological signal and the instrumental noise and independent of the foregrounds.  

\section{The bias on parameters}
\label{apx:bias_par}

Let us assume we have a $n$-sized data vector $\vec{{\bf X}}$ (or a $n\times1$ matrix) with covariance $\Sigma$, with a Gaussian likelihood given some parameters $m$ parameters array ${\bf \vartheta}$ is 
\be
{\cal L} ({\bf X} | \vartheta )= \frac1{\l2\pi \det \Sigma\r^{n/2}}\ e^{-\frac12 ({\bf X}-\bar {\bf X})^{T}\Sigma^{-1}({\bf X}-\bar {\bf X})}\,,
\ee
with $\bar {\bf X}$ the values of ${\bf X}$ that maximise the likelihood. Note that although omitted both $\bar {\bf X}$ and $\Sigma$ depend on $\vartheta$. Let us also assume that the posterior probability of the parameters given the data is Gaussian and expressed as 
\be
{\cal P} (\vartheta | {\bf X} )= \frac1{\l2\pi \det {\bf F}^{-1}\r^{m/2}}\ e^{-\frac12 ({\bf \vartheta}-\bar {\bf \vartheta})^{T}{\bf F}({\bf \vartheta}-\bar {\bf \vartheta})}\,,
\ee
where ${\bf F}$ is the inverse of the covariance of the parameters we which to measure. One can also see that 
\be
{\bf F}_{\vartheta_i\vartheta_j}=-\frac{\partial^2 \ln {\cal P} (\vartheta | {\bf X} )}{\partial \vartheta_i \partial \vartheta_j}\Bigg|_{\vartheta_i=\bar \vartheta_i,\vartheta_j=\bar \vartheta_j}
\ee
In fact even if we had not assumed the Gaussian case, one could still define the information matrix assuming that a maximum of the posterior exists at $\bar {\bf \vartheta}$. Bayes theorem states that
\be
{\rm Posterior\ probability} \propto {\rm Likelihood} \times {\rm Prior} \,.
\ee
Then, in the absence of priors, one can write the information matrix as it is commonly found in the literature
\be
{\bf F}_{\vartheta_i\vartheta_j}=-\frac{\partial^2 \ln {\cal L} ({\bf X} | \vartheta )}{\partial \vartheta_i \partial \vartheta_j}\Bigg|_{\vartheta_i=\bar \vartheta_i,\vartheta_j=\bar \vartheta_j}\,.
\ee
For this paper our data points are the observed angular power spectra, i.e., ${\bf X}=C_\ell(z_i,z_j)$. From this one gets Eq. (\ref{eq:finfo}) assuming that the 4 point function $\Gamma$ is weakly dependent on the cosmological parameters. Note that all assumes the existence of a local maximum of the posterior ($\partial_{\vartheta} {\cal P}|_{\bar\vartheta}=0$), not necessarily that the distributions are Gaussian. 

Now let us assume that we have a model with parameters $\psi$ and a subset of those, $\varphi$ that we will fix while the remaining $\vartheta$ parameters we wish to fit. We know that at the maximum
\be
\partial_{\psi_m} \ln {\cal L}|_{\bar \psi_m} = 0\,,
\ee
therefore at
\be \label{eq:maxtheta}
\partial_{\vartheta_j} \ln {\cal L(\vartheta| \bar \varphi)}|_{\bar \vartheta_j} = 0\,.
\ee
If one expands the likelihood around the maximum 
\bea
 \ln {\cal L(\vartheta|  \varphi)}&=&  \ln {\cal L(\bar \vartheta | \bar \varphi)} + \delta \vartheta_i~ \partial_{\vartheta_i}\l \ln {\cal L( \vartheta |  \varphi)}\r|_{\bar \vartheta, \bar\varphi} \\
&&+\delta \varphi_m~ \partial_{\varphi_m}\l \ln {\cal L(  \vartheta | \varphi)}\r|_{\bar \vartheta, \bar\varphi}
\eea
Then taking the derivative with respect to $\vartheta_j$, and using Eq. \ref{eq:maxtheta} we get
\bea
0&=&\delta \vartheta_i~ \partial_{\vartheta_j}\partial_{\vartheta_i}\l \ln {\cal L( \vartheta |  \varphi)}\r|_{\bar \vartheta, \bar\varphi} +\nn\\
&&+\delta \varphi_m~ \partial_{\vartheta_j} \partial_{\varphi_m}\l \ln {\cal L(  \vartheta | \varphi)}\r|_{\bar \vartheta, \bar\varphi}\,,\nn\\
&=&\delta \vartheta_i\ {\bf H}_{\vartheta_i\vartheta_j} + \delta \varphi_m~ {\bf F}_{\varphi_m\vartheta_j} \,,
\eea
where we defined the information matrix of the smaller set of parameters $\{\vartheta\}$ as ${\bf H}$. In fact ${\bf H}$ is a subset of ${\bf F}$. Then we have that the bias on the best fit model by fixing a larger model with the wrong parameters is given by
\be
\delta \vartheta_i=- \delta \varphi_m~ {\bf F}_{\varphi_m\vartheta_j} {\bf H}^{-1}_{\vartheta_j\vartheta_i}\,.
\ee

\bsp 
\label{lastpage}
\end{document}